# DESIGN, PROTOTYPING AND TESTING OF A COMPACT SUPERCONDUCTING DOUBLE QUARTER WAVE CRAB CAVITY

Binping Xiao[1], Luís Alberty[3], Sergey Belomestnykh[1,2], Ilan Ben-Zvi[1,2], Rama Calaga[3], Chris Cullen[1], Ofelia Capatina[3], Lee Hammons[1], Zenghai Li[4], Carlos Marques[1], John Skaritka[1], Silvia Verdú-Andres[1], Qiong Wu[1]

[1] *Brookhaven National Laboratory, Upton, New York 11973-5000, USA*
[2] *Stony Brook University, Stony Brook, New York 11794, USA*
[3] *European Organization for Nuclear Research (CERN), CH-1211 Geneva 23, Switzerland*
[4] *SLAC national accelerator Laboratory, Menlo Park, CA 94025, USA*

A novel design of superconducting Crab Cavity was proposed and designed at Brookhaven National Laboratory. The new cavity shape is a Double Quarter Wave or DQWCC. After fabrication and surface treatments, the niobium proof-of-principle cavity was cryogenically tested in a vertical cryostat. The cavity is extremely compact yet has a low frequency of 400 MHz, an essential property for service for the Large Hadron Collider luminosity upgrade. The electromagnetic properties of the cavity are also well matched for this demanding task. The demonstrated deflecting voltage of 4.6 MV is well above the requirement for a crab cavity in the future High Luminosity LHC of 3.34 MV. In this paper we present the design, prototyping and test results of the DQWCC.

## I. Introduction

Devices to change the transverse motion of particle beams are widely used in accelerators. The deflecting systems provide a kick to the center of mass of particle bunches whereas crabbing systems kick head and tail of the bunch in opposite directions with a net zero kick to its center of mass. Radio-frequency deflection systems are used as beam choppers [1-3] and beam separators [4-8]. Examples of applications employing crabbing systems are: ultra-short beam pulse generators [9-11], measurement of longitudinal and transverse phase space and bunch length [12-14], ultra-short femtosecond diagnostics [15-18], and "crab crossing" for head-on collision in particle colliders [19]. The concept of crab crossing, which utilizes transverse RF field excited in a resonator called crab cavity, was proposed by Palmer [20], and first implemented in the electron-positron collider KEKB [21]. In collider applications, where a large transverse kick has to be delivered within a limited space available near an interaction point (IP), superconducting RF (SRF) crab cavities are employed.

LHC at CERN is an energy frontier machine for high-energy particle physics, with energy up to 7 TeV per nucleon. The High Luminosity upgrade of the LHC (HL-LHC) includes installation of new Interaction Region (IR) quadrupoles [22] to reduce beam sizes at the interaction points. This demands larger crossing angles to alleviate the beam-beam interaction effects. An implementation of the crab crossing technique [23-29] can be used to modify the angle at which bunches collide, and hence would complement the IR quadrupole upgrade to maximize the LHC luminosity.

The limited space available in the surroundings of the ATLAS and CMS detectors, respectively, at IP1 and IP5, limits dimensions of the crab cavities. An upper frequency limit is set to minimize nonlinearities of the crabbing kick given to bunches. This resulted in a relatively low resonant frequency of the HL-LHC crab cavities at 400 MHz. Currently there are three compact crab cavity designs as potential candidates [24-29] and the double quarter wave crab cavity (DQWCC), the subject of this article, is one of them.

The LHC crab cavity project is divided into three phases. The first phase aims to validate cryogenic performance of proof-of-principle (PoP) cavities. The cavities must demonstrate a deflecting voltage of 3.34 MV per cavity. The second phase comprises the design and prototyping of a pair of fully dressed cavities that will be tested with beam in the Super Proton Synchrotron (SPS) at CERN in 2017-2018. The third phase is dedicated to the design, manufacturing and installation of the crab cavities in LHC, with 4 cavities per beam at each side of the IP, 32 cavities in total plus spare cavities.

Quarter wave resonators (QWRs) are widely used in particle accelerators. The first superconducting QWR for ion acceleration was proposed by Ben-Zvi and Brennan in 1983 [30]. Since then, QWRs have been successfully employed in many low $\beta$ linacs, as it is evident from Ref. [31-35]. In recent years, this cavities found new applications in high $\beta$ accelerators, such as described in [36-40]. The QWR was first proposed as a deflecting/crabbing cavity by Ben-Zvi [41, 42]. The design eventually evolved into a symmetric double quarter wave structure [26]. Comparing to the conventional squashed elliptical cavity [21], the DQWCC is very compact and its crabbing mode is the lowest resonant frequency $f_0$ of the cavity. No lower order mode (LOM) or same order mode (SOM) exists in this structure, and the first higher order mode (HOM) is well separated from $f_0$ by an amount of about $f_0/2$ due to the large capacitance. In this article we describe design, fabrication and test results of the PoP DQWCC, built as part of the HL-LHC upgrade.

## II. Cavity Design

A. Cavity RF design

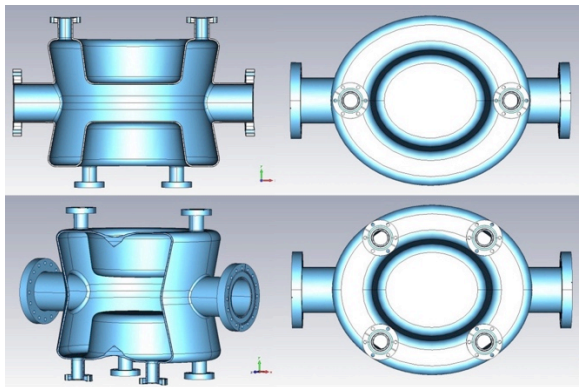

Figure 1. Geometry of the DQWCC.

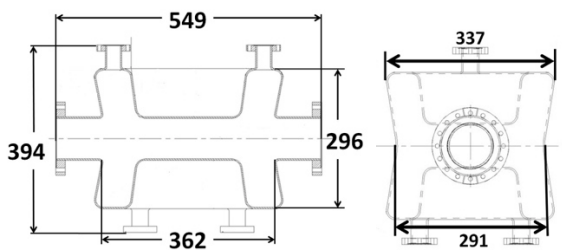

Figure 2. Envelope (in mm) of the DQWCC.

A similar design of a normal-conducting deflector for low velocity particles was proposed by Krawczyk [43]. The DQWCC is the first SRF cavity design of this kind, in particular for ultra-relativistic applications. The cavity, shown in Figure 1, can be considered as two quarter-wave resonators sharing a load capacitor, thus dubbed as a double quarter wave cavity. At the fundamental mode, there is a transverse electric field between two plates of the capacitor, which offers the crabbing voltage when the beam passes at an appropriate phase.

An envelope of the cavity is shown in Figure 2. The envelope, in particular the cavity waist and height, along with all necessary ports, is optimized to fit the size limit of IP1 and IP5 in LHC for both crabbing directions, as depicted in Figure 3. Figure 4 illustrates the field distributions inside the DQWCC. The magnetic field of the fundamental mode is concentrated in the coaxial area. The waist of the cavity provides clearance to the second beam pipe for vertical kick. The height of the cavity is maintained short enough to give clearance to the second beam pipe for horizontal kick, shown in Figure 3b.

Comparing to the former quarter wave version [25], the DQWCC is optimized to cancel the on-axis accelerating (longitudinal) field and to reduce the overall nonlinearity of the deflecting voltage as a function of offset [26] from the symmetry introduced by changing from quarter wave to double quarter wave resonator. The field nonlinearity, caused by the axial asymmetry of the cavity, is characterized by its multipolar expansion coefficients as described in Ref. [44, 45], with $b_3 = 1070$, $b_5 = -1\times10^5$, $b_7 = 7\times10^6$. The other components are zero for the notation used in [44]. These numbers are well below the requirements specified in [45].

The key RF parameters of the DQWCC are listed in Table 1. Besides the two beam pipe ports, 6 extra ports are located on the top and bottom, with two on one top and four on the bottom, as shown in Figure 1. These ports are designated for a fundamental power coupler (FPC), an RF pickup (PU) probe, and higher order mode couplers.

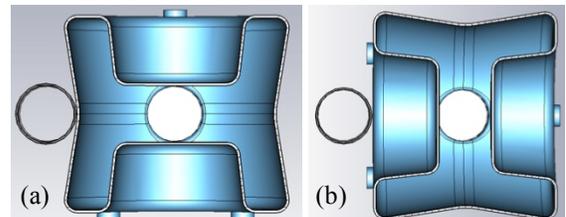

Figure 3. DQWCC with adjacent beam pipes for: (a) vertical kick scheme and (b) horizontal kick scheme.

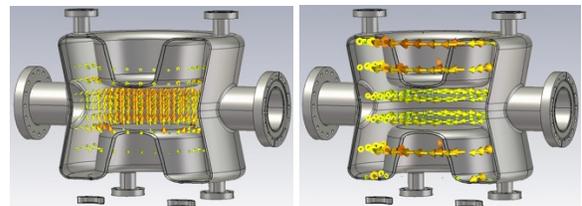

Figure 4. Electric (left) and magnetic (right) field distributions inside the DQWCC.

Table 1. Key RF parameters of the double quarter wave crab cavity.

| | |
|---|---|
| Fundamental mode frequency $f_0$ [MHz] | 400 |
| Nearest HOM frequency $f_1$ [MHz] | 579 |
| Deflection voltage $V_t$ [MV] | 3.34 |
| $R_t/Q$ (fundamental mode) [Ω] | 406 |
| Geometry factor [Ω] | 85 |
| Peak surface electric field $E_{peak}$ [MV/m] | 49.0 |
| Peak surface magnetic field $B_{peak}$ [mT] | 80.0 |
| Residual accelerating voltage $V_{acc}$ [kV] | 1.6 |
| Stored energy [J] | 10.9 |

### B. Cavity mechanical design

Based on the safety requirements at CERN [46, 47], with a relief valve attached to the cryogenic test dewar set at 1.5 bar, the cavity has to pass a proof test with the maximum pressure of 2 bar outside the cavity and vacuum inside the cavity. The safety factor applied to the relevant material properties is 1.05, being the maximum allowable stress given by the yield strength divided by the safety factor. The failure mode considered for the presented strength assessment is gross plastic deformation: the maximum allowable stress of high RRR niobium, the material that the cavity is made of, is 70 MPa [48, 49]. Software ANSYS$^{TM}$ was used for stress and deformation analysis, with a 2 mm mesh size for all simulations. Gravity was considered during the simulations. The cavity is positioned in a dewar with its beam axis in vertical direction during the test.

The above-mentioned 2 bar pressure difference makes the two capacitive plates to move closer, and the high magnetic field region to deform, generating high stresses in that region. The standard niobium sheets come in different thicknesses varying from 2.8 mm to 4.0 mm. If a 3 mm thick niobium were used, a maximum total deformation of 1.2 mm would be at the center of the capacitive plate and a maximum stress intensity of 150 MPa would be located in the magnetic field area, exceeding the maximum allowable stress of niobium. We chose 4 mm thick niobium sheets to lower the maximum stress intensity on the cavity body, with the beam pipes made of niobium tubes with 2.8 mm wall thickness. In the simulations, a reduction of 0.21 mm in thickness resulting from the cavity surface treatment was accounted for. However, even with thicker walls, the mechanical stress on the cavity is still too high.

The cavity is reinforced and supported by holding the two capacitive plates with a dedicated grade 2 titanium frame made of 2″ wide and 0.5″ thick titanium bars, as shown in Figure 5. Four niobium bars were electron beam welded to the capacitive plates. Then titanium plates with serrated surface were bolted to the niobium bars, with an aid of another titanium bar with serrated surface on the other side of each niobium bar. The properties of niobium and titanium are listed in Table 2, with data taken from Ref. [48] and [50].

Table 2. Properties of niobium (Nb) and titanium (Ti) at room temperature

| | Density [kg/m$^3$] | Modulus [GPa] | Poisson's ratio | Yield strength [MPa] |
|---|---|---|---|---|
| Nb | 8570 | 103 | 0.38 | 75 |
| Ti | 4510 | 102 | 0.34 | >276 |

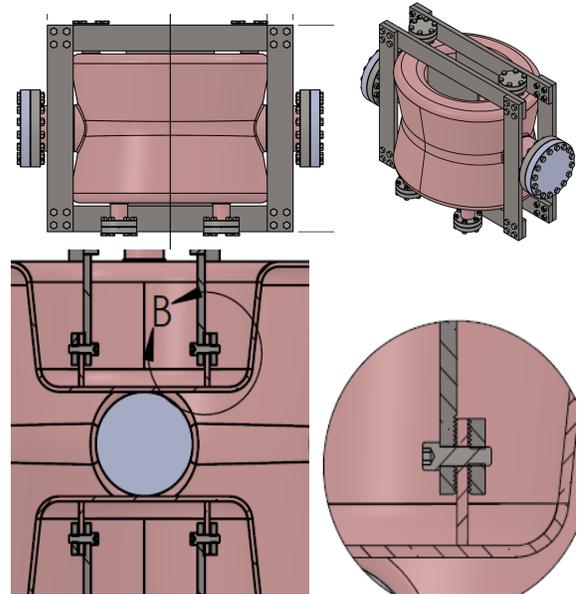

Figure 5. Geometry of the PoP DQWCC with titanium stiffening frame.

The simulation model was modified to be symmetric along the vertical symmetry plane and only one half of the model was used to reduce the calculation time.

The maximum deformation due to a pressure load of 2 bar was around 0.5 mm for the cavity with stiffener, located at the center of the capacitive plates. A maximum stress intensity of 230 MPa was found on the stiffening titanium frame, while the highest stress on the cavity is 67 MPa, below the maximum allowable stress of niobium.

## C. Multipacting

Multipacting is an electron avalanche effect due to resonant multiplication of secondary electrons. It absorbs RF power, hence limiting the power available to excite the fields in a cavity [51], causing the quality factor to drop and limiting the maximum cavity gradient. Multipacting depends on the secondary electron yield of the cavity surface material and the cavity shape. Numerical analysis using the TRACK3P solver from the SLAC ACE3P suite of codes [52] showed that multipacting appears at the waist area of the cavity between the cavity wall and the capacitive plates at a deflecting voltage of about 0.1 MV with an electron impact energy up to 1500 eV, as shown in Figure 6. The number of the electrons with energy above 100 eV indicates that multipacting might happen at low deflecting voltage, and 2~3 MV deflecting voltage.

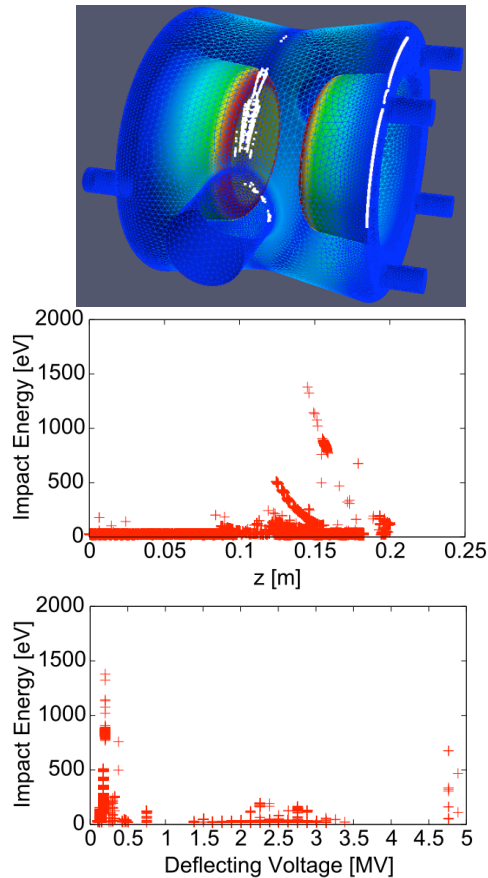

Figure 6. Multipacting studies performed for the DQWCC with TRACK3P. Top: Potential multipacting sites on the cavity geometry are shown as white spots; Middle: Electron impact energy along the vertical direction at 0.1 MV kick, with the maximum value to be 1500 eV; Bottom: Electron impact energy for different deflecting voltage values.

## III. Cavity Fabrication and Measurements at Room Temperature

### A. Cavity fabrication and processing

The cavity body was fabricated by Niowave, Inc. from 4 mm niobium sheets using aluminum dies and then electron beam welded together. Figure 7 shows the cavity body parts before welding. 2.8 mm thick niobium tubes, with 84 mm inner diameter for beam pipes and 28 mm inner diameter for FPC, PU, and HOM ports, were electron beam welded to the cavity body. The CF flanges were brazed to the tubes prior to welding. Machining tolerances were not specified as the resonance frequency was not critical for the PoP cavity.

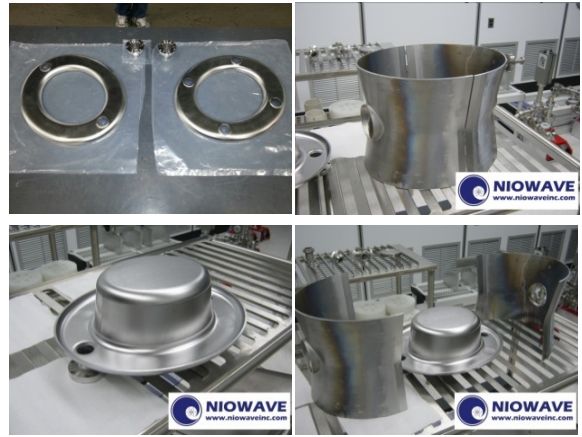

Figure 7. Cavity body fabrication, top-left: top and bottom caps; top-right: outer tube; bottom-left: capacitive plate welded to the inner tube and cap; bottom-right: parts before welding.

After welding, the cavity was chemically treated using 1:1:2 buffered chemical polish (BCP) solution of HF (49% wt), $HNO_3$ (69% wt), and $H_3PO_4$ (85% wt) to etch 150 μm of the inner niobium surface. Once at BNL, the cavity was visually inspected, then leak checked, and finally baked for 10 hour at 600°C in a vacuum oven. After baking the resonant frequency was measured to be 403.3117 MHz. The cavity quality factor at room temperature was 5,400, in agreement with simulation result for niobium with an electric conductivity of $6.2 \times 10^6$ /(Ω·m) at room temperature. The cavity was then shipped back to Niowave for a light BCP (30 μm material removal) and high pressure rinse (HPR) with de-ionized ultra-pure water. Then it was shipped to BNL for assembly and cryogenic test.

After the first cold test, further surface treatment was performed at ANL. First, the cavity underwent an ultrasonic cleaning, which improved significantly the brightness of the cavity outer surface. Then the cavity was BCP etched to remove 40 μm. After that the cavity was rinsed with distilled water, and ultrasonically degreased with a 2% Liquinox solvent, followed by another rinsing in a bath of water. Finally, the cavity went through HPR with deionized water at a pressure of 1200 psi. The cavity was then shipped back to BNL for the second cryogenic test after an additional 24 hour baking at 120°C with cavity inside under vacuum.

### B. Measurements at room temperature

Measurements at room temperature can help to understand how cavity may change during the cavity fabrication and surface treatment. This shall help to better specify the quality control steps for the SPS cavities. The thickness of the cavity was verified using an ultrasonic thickness detector with sensitivity of 0.01 mm and accuracy of 0.03 mm. Starting with 4 mm thick niobium, after one heavy BCP at 150 μm and two light BCP at 30 μm and 40 μm, respectively, the thickness was expected to be 3.78 mm. 52 points were chosen on the outer surface of the cavity. The measured thickness is 3.92 mm, with a standard deviation of 0.08 mm. Thus the cavity thickness is 140±80 μm thicker than expected, which indicates insufficient material removal during chemical etching.

The cavity profile was measured using a laser tracker with an accuracy of 0.2 mm. The flanges of the beam pipes were used as a reference for this measurement. The two capacitive plates of the cavity were found to be tilted 0.21° with respect to the center beam line. The bottom cap on the four-port side is titled 0.29°. The top cap on the two-port side is tilted 0.63°. All tilts are in the same direction.

The fundamental mode field profile was studied using a bead-pulling setup [53]. The measurements were performed along the beam axis, and 35mm vertically away from the axis, in a direction that the bead is moving in parallel to the beam axis and 7mm away from the the two-port top cap, using a Micarta$^{TM}$ dielectric bead (4 mm radius, 1.5 mm thick cylindrical bead) that measures the electric field profile. The field profiles are shown in Figure 8. The measured result with bead pulling along the axis is consistent with the simulation result. The 35 mm vertically off axis shows two peaks, shown in circles in Figure 8, slightly away from the simulation result, explained by the tilting of the two capacitive plates detected in the laser tracker profile measurements.

## IV. Cold Test

### A. Cryogenic test preparation

The cavity was tested in a small vertical test facility (SVTF) at BNL. A 360 Watt 4 K refrigerator and a liquid ring pump are used to reach 1.9 K. The cryogenic system is capable of handling a 100 Watt heat load. A 200 W OPHIR$^{TM}$ RF power amplifier was used to power the cavity. A low level RF phase lock loop was made to lock the RF signal generator frequency to the cavity resonant frequency during the cryogenic test for both CW mode and pulsed mode. Six CERNOX$^{TM}$ temperature sensors were mounted onto the crab cavity, with four of them labeled in Figure 9. Sensors #2 and #3 were placed in high magnetic field regions. Sensors #1 and #4 were placed on the beam pipe flanges to monitor the RF loss on the niobium film coated stainless steel flanges. Sensor #1 is also used to make sure that the whole cavity is covered by liquid helium during the test. A LabVIEW program was developed to record the temperature data every second.

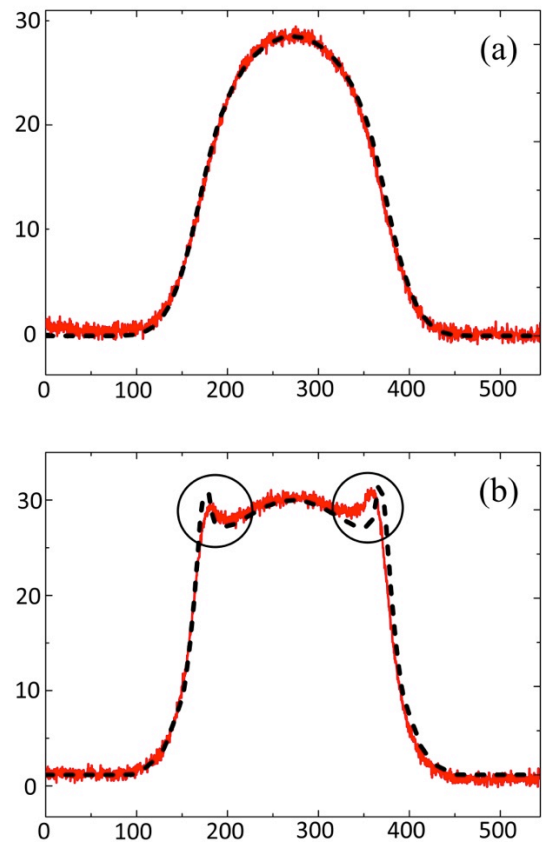

Figure 8. Bead pulling measurement using Micarta$^{TM}$ bead, (a) along the beam axis; (b) 35mm vertically away from the beam axis. $x$ axis represents the position along the beam axis from one beam port to the other one in mm, $y$ axis represents the phase difference in degrees. Black dots are the simulation field results and red curves are the measurement results.

For the FPC probe, a hook shaped coupler was chosen, with its configuration shown on the right of

Figure 9. The FPC was set up to provide $Q_{ext}$ of $2.9\times10^8$ to $5.4\times10^{10}$ with 20 mm travel. Dependence of the external Q factor of the hook position is shown in Figure 10. The nominal penetration of the FPC was set to be 21.2 mm away from the cavity inner surface, as indicated in Figure 9. The corresponding $Q_{ext}$ is $4.0\times10^9$. The motion of the FPC is controlled by a stepper motor mounted on the dewar's top-plate and connected to the FPC via a gear box, a drive shaft and an isosceles linkage to transfer vertical to horizontal motion.

A copper antenna is used as an RF pickup probe, shown in the left top corner of Figure 9, with its position set to be 20.7 mm away from the cavity inner surface, corresponding to $Q_{ext}$ of $1.0\times10^{11}$. The assembling error is controlled to less than 0.5 mm, corresponding to $Q_{ext}$ uncertainty of $8.8\times10^{10}$ to $1.2\times10^{11}$.

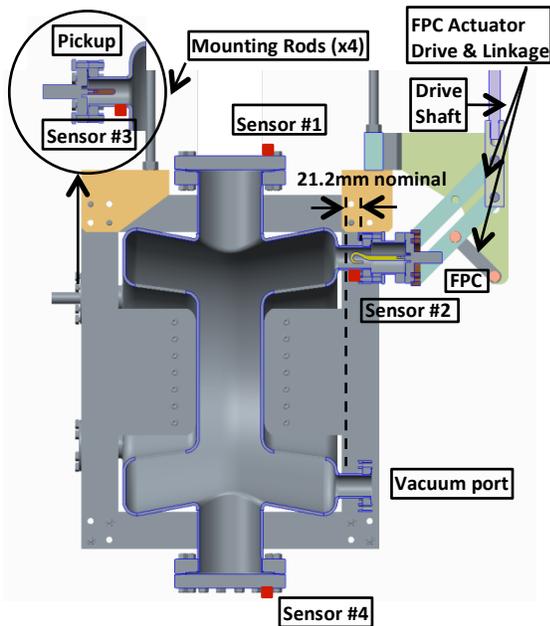

Figure 9. Configuration of couplers and temperature sensors.

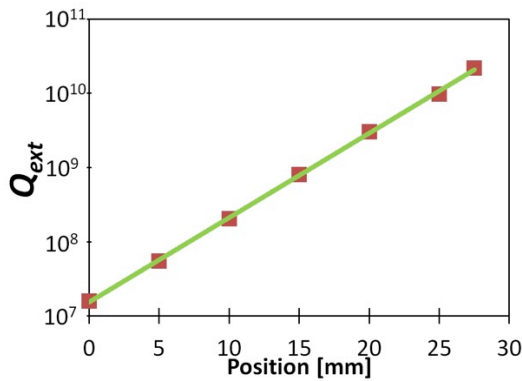

Figure 10. $Q_{ext}$ vs. FPC hook position.

Two beam pipe ports were sealed with niobium-plated flanges. With an assumed surface resistance of the niobium thin film to be 20 nΩ during the cold test, the quality factor caused by the power loss on these two flanges is $3.0\times10^{15}$. Three small ports that are not used during this test were sealed with copper disk gaskets with a small venting hole in the gasket center. The quality factor corresponding to the power loss on these gaskets is $1.1\times10^{11}$. The power loss on the FPC copper hook in nominal position with a stainless steel feed-through gives an external quality factor of $4.5\times10^{10}$, and for the pickup copper antenna with stainless steel feed-through, it is $2.9\times10^{11}$.

Residual magnetic field around the cavity in the dewar was measured to be 1.2 µT, close to the shielding requirement in the LHC tunnel of 1 µT. The residual magnetic field induced residual surface resistance is 2.3 nΩ [51].

### B. Cryogenic testing and analysis

The first cryogenic test was carried out in June of 2013. Figure 11 shows rigging the cavity into the dewar, and the dewar with cavity inside the SVTF. Multipacting was encountered in the deflecting voltage range between 0.07 and 0.16 MV, consistent with the multipacting analysis presented in Figure 6. It was quickly conditioned during the first field ramp up and never observed afterwards. $Q_0$ was limited to below $3\times10^8$ even at low field level, and did not improve after cooling down from 4.3 K to 1.9 K. Heating on the beam pipe flanges (Sensors #1 and #4 in Figure 9) was observed in correlation with the $Q_0$ decrease, which indicated poor quality of the niobium coating. The cavity reached a maximum deflecting voltage of only 1.34 MV, limited by the RF power amplifier.

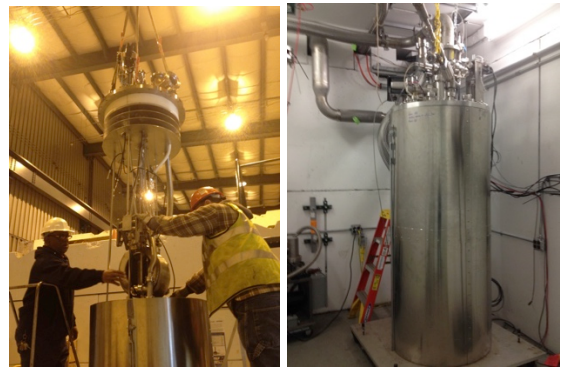

Figure 11. Left: rigging the cavity into the dewar, right: the dewar with cavity inside the SVTF.

The second cryogenic test was performed in November of 2013 after additional treatment of the cavity at ANL as described above. With 1.9 K bath temperature, as the RF power was increased, $Q_0$ started degrading at 2.0 MV, associated with high radiation due to field emission, with its peak reached 864 mR/h at 3.0 MV. The $Q_0$ recovered after about 30 minutes of RF conditioning, as shown in Figure 12. The radiation level went down below 15 mR/h. The highest voltage reached during the test was 4.6 MV in CW mode, limited by quench, also shown in Figure 12. This conditioning might also be associated with multipacting, since the simulations showed a multipacting band at 2~3 MV.

During the test, we observed temperature rise on both beam pipe flanges (Sensors #1 and #4 in Figure 9) and pickup port blending area (Sensors #3 in Figure 9).

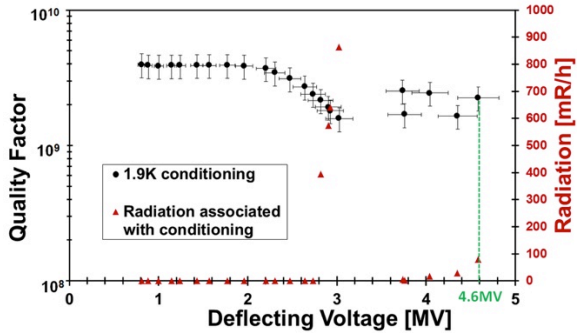

Figure 12. Conditioning of the DQWCC at the helium bath temperature of 1.9 K. Black dots represent the quality factor and red triangles correspond to the associated radiation level. The cavity was RF conditioned at the deflecting voltage of 3.0 MV for about half an hour. There was high radiation associated with field emission. After conditioning, the deflecting voltage jumped to 3.8 MV and eventually reached 4.6 MV.

Figure 13 shows the quality factor measurement results after conditioning. At 4.3 K liquid helium bath temperature, the measured $Q_0$ reduced from $1.2 \times 10^9$ to $4.4 \times 10^8$ with the deflecting voltage increasing from 0.1 MV to 2.6 MV. The power decreasing measurement, however, did not follow the power increasing measurement curve due to the residual stored heat from high power. The measured $Q_0$ at 1.9 K is in the range from $3.0 \times 10^9$ to $4.0 \times 10^9$. In CW mode, temperature of the beam pipe flanges raised and caused degradation of $Q_0$ starting at around 3.3 MV. The plots in Figure 13 illustrate thermal hysteresis due to heating of the beam pipe flanges. The temperature increase was observed at the pickup port blending area as well.

In an attempt to reduce the heating on the beam pipe flanges, the cavity was further tested in pulsed mode. The highest voltage reached was 4.5 MV, limited by quench, consistent with the test results in CW mode at 4.6 MV. The cavity quenches at a peak magnetic field of about 110.2 mT, and peak electric field of about 67.5 MV/m.

The surface resistance at low field is 60.0 nΩ at 4.3 K and 22.1 nΩ at 1.9 K. The measurement results are fitted using the following expression [51, 54]:

$$R_s[n\Omega] = \frac{A}{T}ln\left(\frac{4kT}{hf}\right)e^{(-\frac{\Delta}{kT})} + R_{res} \qquad (1)$$

The fitting parameters are $R_{res} = 21$ nΩ, $\Delta/kT_c = 1.87$ and $A = 1212.77$ nΩ·K. Figure 14 shows comparison of the measured data points and the fitting curve. We suspect that high residual resistance is partly associated with losses in the thin niobium coating of stainless steel flanges on the beam pipes. The temperature rise on the flanges indicates that the quality of the thin film was not well controlled. Another possible reason for high residual resistance is insufficient etching during BCP, which is corroborated by the cavity wall thickness measurement.

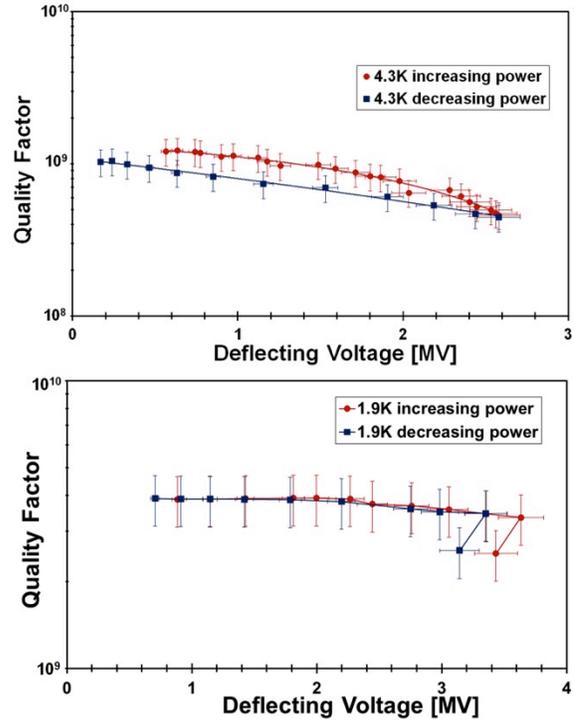

Figure 13. Cavity quality factor at 4.3 K (top) and 1.9 K (bottom) bath temperatures in CW mode. The data illustrate thermal hysteresis associated with heating of the beam pipe flanges. The red curves were taken with RF power ramping up and the black curves were taken with RF power ramping down.

The Lorentz force detuning is a cavity shape deformation effect brought by radiation pressure, with magnetic field causing pressure outward the cavity and electric field causing pressure inward. It causes frequency shift with the cavity electromagnetic field. The Lorentz force detuning coefficient of the DQWCC is -206 Hz/MV$^2$, calculated from the CW measurement at 1.9 K. Sensitivity of the cavity frequency to the helium bath pressure was measured during the cavity cool down. The resonance frequency changed from 403.667 MHz at 4.3 K to 403.930 MHz at 1.9 K. The pressure sensitivity is -336 Hz/Torr.

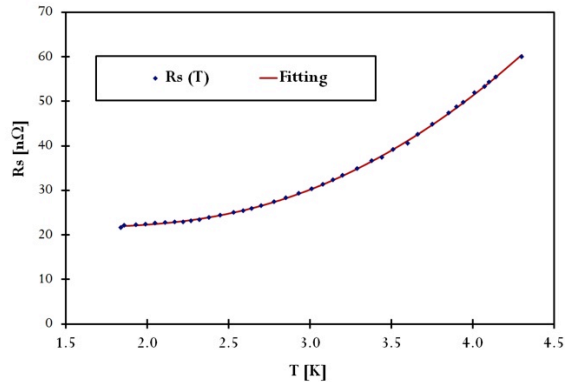

Figure 14. Surface resistance versus cavity temperature during the cavity cool down from 4.3 K to 1.9 K. The data points are taken at peak RF fields lower than 22.5 mT. Dots are measured data points; the red curve is fitting using equation (1).

## V. Conclusion

A double quarter wave resonator was proposed as a compact crab cavity for the LHC luminosity upgrade. A PoP cavity has been designed and fabricated to validate that the cavity of such complicated shape can deliver deflecting voltage over required 3.34 MV. A special stiffening frame was designed and implemented for the PoP cavity to relieve mechanical stresses on the cavity and bring it in compliance with the CERN safety code. The cavity thickness is measured using an ultrasonic thickness detector and it is 140 µm thicker than expected, which indicates insufficient material removal during chemical etching. The cavity profile measurement using a laser tracker reveals that top and bottom caps and two capacitive plates are tilted in the same direction with respect to the center beam line. Bead-pulling measurement using dielectric bead moving along the beam axis is consistent with the simulated electric field profile. Similar measurement with bead moving 35 mm vertically off axis showed two peaks that are slightly away from the simulation result, explained by the tilting of the two capacitive plates detected in the laser tracker profile measurements. A multipacting zone was found at a deflecting voltage of approximately 0.1 MV during the first cold test, as predicted by numerical analysis, and was easily conditioned. The cavity did not show any multipacting in this voltage range during the second cold test. During the cryogenic tests conducted at BNL the cavity achieved a deflecting voltage of 4.6 MV, with a quality factor higher than $3.0 \times 10^9$. Possible reasons for the relatively low quality factor are associated with i) an insufficient BCP etching, and ii) poor quality of niobium coating on the stainless steel flanges for beam pipes.

With the PoP DQWCC cavity exceeding design voltage by a healthy margin, we are proceeding with the design and fabrication of two fully dressed prototype cavities. After verification of the cavities' performance in vertical tests, they will be installed in SPS testing with beam as the next step toward implementation of the crab cavity scheme in the HL-LHC.

The DQW design can be applied to other particle accelerators requiring deflecting or crab cavities, in particular to the future electro-hadron collider eRHIC at BNL [55]. In eRHIC, the 5 to 8 cm long ion bunches will collide with electrons at a crossing angle of 4 mrad. This would require a low crabbing frequency $f_0$, currently set at 225 MHz. DQWCC is well suited for such low frequency application with its reasonable geometrical size [42, 56].

## ACKNOWLEDGEMENT

The authors would like to acknowledge Niowave, Inc for fabrication of this cavity. The authors would like to thank S. Gerbick, M. P. Kelly, R. C. Murphy, P. N. Ostroumov and T. C. Reid at ANL for the surface treatment, and D. Beavis, P. P. Cirnigliaro, C. Degen, H. Dorr, A. Etkin, R. C. Karol, R. Kellermann, E. T. Lessard, G. T. McIntyre, J. Moore, S. P. Pontieri, R. Porqueddu, T. Seda, L. P. Snydstrup, T. Tallerico, R. Than and J. E. Tuozzolo at BNL for help with the cryogenic setup and test. Work partly supported by the US DOE through Brookhaven Science Associates, LLC under contract No. DE-AC02-98CH10886 with the US LHC Accelerator Research Porgram (LARP), and by the EU FP7 HiLumi LHC grant agreement No. 284404. This research used resources of the National Energy Research Scientific Computing Center (NERSC), which is supported by the US DOE under contract No. DE-AC02-05CH11231.

## REFERENCES

[1] S. Fu, T. Kato, Nucl. Instrum. Meth. A **440**, 296 (2000). http://dx.doi.org/10.1016/S0168-9002(99)00946-8.
[2] A. Aleksandrov, in *Proceedings of the XXV Linear Accelerator Conference, Tsukuba, Japan,*

2010, p. 689. http://accelconf.web.cern.ch/accelconf/LINAC2010/papers/we104.pdf.
[3] T. Maruta, M. Ikegami, Nucl. Instrum. Meth. A **728**, 126 (2013). http://dx.doi.org/10.1016/j.nima.2013.06.046.
[4] H. Hahn, H. J. Halama, Rev. Sci. Instrum. **36**, 1788, (1965). http://dx.doi.org/10.1063/1.1719466.
[5] A. Citron, G. Dammertz, M. Grundner, L. Husson, R. Lehm, H. Lengler, Nucl. Instrum. Meth. **155**, 93 (1978). http://dx.doi.org/10.1016/0029-554X(78)90190-8.
[6] C. Leemann, C. G. Yao, in *Proceedings of the Linear Accelerator Conference 1990, Albuquerque, New Mexico, USA, 1990,* p. 232. http://accelconf.web.cern.ch/accelconf/l90/papers/mo465.pdf.
[7] R. Tomas, Phys. Rev. Spec. Top. Accel. Beams **13**, 014801, (2010). http://dx.doi.org/10.1103/PhysRevSTAB.13.014801.
[8] M. Aicheler, P. Burrows, M. Draper, *et al.*, Report CERN-2012-007, 2012 (unpublished). http://dx.doi.org/10.5170/CERN-2012-007.
[9] M. Katoh, Jpn. J. Appl. Phys. **38**, 547 (1999). http://iopscience.iop.org/1347-4065/38/5A/L547.
[10] P. Emma, R. Iverson, P. Krejcik, P. Raimondi, J. Safranek, in *Proceedings of the 2001 Particle Accelerator Conference, Chicago, Illinois, USA, 2001,* p. 4038. http://accelconf.web.cern.ch/AccelConf/p01/PAPERS/FPAH165.PDF.
[11] Z. Huang, A. Brachmann, F.-J. Decker, *et al.*, Phys. Rev. Spec. Top. Accel. Beams **13**, 020703, (2010). http://dx.doi.org/10.1103/PhysRevSTAB.13.020703.
[12] P. Emma, J. Frisch, P. Krejcik, Report LCLS-TN-00-12, 2000 (unpublished). http://www-ssrl.slac.stanford.edu/lcls/technotes/lcls-tn-00-12.pdf.
[13] D. Alesini, G. D. Pirro, L. Ficcadenti, *et al.*, Nucl. Instrum. Meth. A **568**, 488 (2006). http://dx.doi.org/10.1016/j.nima.2006.07.050.
[14] A. Falone, H. Fitze, R. Ischebeck, *et al.*, in *Proceedings of the 23rd Particle Accelerator Conference, Vancouver, BC, Canada, 2009,* p. 2012. http://accelconf.web.cern.ch/accelconf/pac2009/papers/we5pfp012.pdf.
[15] H. Tomizawa, H. Hanaki, T. Ishikawa, in *Proceedings of 29th International Free Electron Laser Conference, Novosibirsk, Russia, 2007,* p. 472. http://accelconf.web.cern.ch/AccelConf/f07/PAPERS/WEPPH053.PDF.
[16] Y. Ding, C. Behrens, P. Emma, *et al.*, Phys. Rev. Spec. Top. Accel. Beams **14**, 120701, (2011). http://dx.doi.org/10.1103/PhysRevSTAB.14.120701.
[17] R. K. Li, P. Musumeci, H. A. Bender, N. S. Wilcox, M. Wu, J. Appl. Phys. **110**, 074512, (2011). http://dx.doi.org/10.1063/1.3646465.
[18] V. A. Dolgashev, J. Wang, in *AIP Conference Proceedings, 2012,* p. 682. http://dx.doi.org/10.1063/1.4773780.
[19] K. Akai, M. Tigner, Deflecting and Crab Cavities, In A. W. Chao, K. H. Mess, M. Tigner, F. Zimmermann, (Eds.), *Handbook of Accelerator Physics and Engineering*. 3rd ed World Scientific., (2006). p. 617.
[20] R. B. Palmer, Report SLAC-PUB-4707, 1988 (unpublished). http://www.slac.stanford.edu/cgi-wrap/getdoc/slac-pub-4707.pdf.
[21] T. Abe, K. Akai, M. Akemoto, *et al.*, in *Proceedings of 22nd Particle Accelerator Conference, Albuquerque, New Mexico, USA, 2007,* p. 27. http://accelconf.web.cern.ch/AccelConf/p07/PAPERS/MOZAKI01.PDF.
[22] L. Rossi, S. Stavrev, A. Szeberenyi, Report CERN-ACC-2013-022, 2013 (unpublished).
[23] Z. Li, L. Xiao, C. Ng, T. Markiewicz, in *Proceedings of the 1st International Particle Accelerator Conference, Kyoto, Japan, 2010,* p. 504. http://accelconf.web.cern.ch/AccelConf/IPAC10/papers/mopec022.pdf.
[24] R. Calaga, in *Proceedings of 15th International Conference on RF Superconductivity, Chicago, IL USA, 2011,* p. 988. http://accelconf.web.cern.ch/accelconf/SRF2011/papers/friob05.pdf.
[25] R. Calaga, S. Belomestnykh, I. Ben-Zvi, Q. Wu, in *Proceedings of International Particle conference 2012, New Orleans, Louisiana, 2012,* p. 2260. http://accelconf.web.cern.ch/accelconf/IPAC2012/papers/weppc027.pdf.
[26] R. Calaga, S. Belomestnykh, I. Ben-Zvi, J. Skaritka, Q. Wu, B. P. Xiao, in *Proceedings of the 4th International Particle Accelerator Conference, Shanghai, China, 2013,* p. 2408. http://accelconf.web.cern.ch/accelconf/IPAC2013/papers/wepwo047.pdf.
[27] S. U. D. Silva, J. R. Delayen, Phys. Rev. Spec. Top. Accel. Beams **16**, 082001, (2013). http://dx.doi.org/10.1103/PhysRevSTAB.16.082001

[28] B. P. Xiao, S. Belomestnykh, I. Ben-Zvi, *et al.*, in *Proceedings of 16th International Conference on RF Superconductivity, Paris, France, 2013,* p. 1002. http://accelconf.web.cern.ch/AccelConf/SRF2013/papers/thp043.pdf.
[29] G. Burt, B. Hall, C. Lingwood, *et al.*, in *Proceedings of the 4th International Particle Accelerator Conference, Shanghai, China, 2013,* p. 2420. http://accelconf.web.cern.ch/accelconf/IPAC2013/papers/wepwo051.pdf.
[30] I. Ben-Zvi, J. M. Brennan, Nucl. Instrum. Meth. 212, 7 (1983). http://dx.doi.org/10.1016/0167-5087(83)90678-6.
[31] W. Hartung, J. Bierwagen, S. Bricker, *et al.*, in *XXIV Linear Accelerator Conference, Victoria, BC, Canada, 2008,* p. 854. http://accelconf.web.cern.ch/AccelConf/LINAC08/papers/thp033.pdf.
[32] G. Olry, S. Bousson, T. Junquera, *et al.*, in *2006 Linear Accelerator Conference, Knoxville, Tennessee, USA, 2006,* p. 698. http://accelconf.web.cern.ch/accelconf/l06/PAPERS/THP052.PDF.
[33] A. Roy, in *13th International workshop on RF Superconductivity, Beijing, China, 2007,* p. 24. http://accelconf.web.cern.ch/accelconf/srf2007/PAPERS/MO303.pdf.
[34] R. E. Laxdal, K. Fong, M. Laverty, *et al.*, Physica C **441**, 13 (2006). http://www.sciencedirect.com/science/article/pii/S0921453406001456#.
[35] M. P. Kelly, J. D. Fuerst, S. Gerbick, *et al.*, in *XXIV Linear Accelerator Conference, Victoria, British Columbia, Canada, 2008,* p. 836. http://accelconf.web.cern.ch/accelconf/LINAC08/papers/thp025.pdf.
[36] C. H. Boulware, T. L. Grimm, I. Ben-Zvi, S. Belomestnykh, in *International Particle Accelerator Conference 2012, New Orleans, Louisiana, USA, 2012,* p. 2405. http://accelconf.web.cern.ch/AccelConf/IPAC2012/papers/weppc083.pdf.
[37] Q. Wu, S. Belomestnykh, I. Ben-Zvi, *et al.*, in *16th International Conference on RF Superconductivity, Cite INternationale Universitaire, Paris, France, 2013,* p. 969. http://accelconf.web.cern.ch/AccelConf/SRF2013/papers/thp031.pdf.
[38] S. Belomestnykh, I. Ben-Zvi, C. H. Boulware, *et al.*, in *Particle Accelerator Conference 2011, New York, NY, USA, 2011,* p. 898. http://accelconf.web.cern.ch/AccelConf/PAC2011/papers/tup051.pdf.
[39] J. R. Harris, K. L. Ferguson, J. W. Lewellen, *et al.*, Phys. Rev. Spec. Top. Accel. Beams **14**, 053501, (2011). http://dx.doi.org/10.1103/PhysRevSTAB.14.053501.
[40] J. Bisognano, R. A. Bosch, D. Eisert, *et al.*, in *Particle Accelerator Conference 2011, New York, NY, USA, 2011,* p. 2444. http://accelconf.web.cern.ch/AccelConf/PAC2011/papers/thp176.pdf.
[41] I. Ben-Zvi. *4th LHC Crab Cavity Workshop*; *December 15-17, 2010*; *CERN, Geneva, Switzerland* 2010 (unpublished), http://indico.cern.ch/event/100672/session/1/contribution/52/material/slides/1.pdf.
[42] I. Ben-Zvi, in *Proceedings of 15th International Conference on RF Superconductivity, Chicago, IL USA, 2011,* p. 637. http://accelconf.web.cern.ch/AccelConf/SRF2011/papers/thioa04.pdf.
[43] F. L. Krawczyk, in *Proceedings of the 1995 Particle Accelerator Conference, Dallas, Texas, USA, 1995,* p. 2361. http://accelconf.web.cern.ch/AccelConf/p95/ARTICLES/MPC/MPC22.PDF.
[44] J. B. García, R. Calaga, R. D. Maria, M. Giovannozzi, A. Grudiev, R. Tomás, in *Proceedings of International Particle Accelerator Conference 2012, New Orleans, Louisiana, USA, 2012,* p. 1873. http://accelconf.web.cern.ch/accelconf/IPAC2012/papers/tuppr027.pdf.
[45] P. Baudrenghien, K. Brodzinski, R. Calaga, *et al.*, Report CERN-ACC-NOTE-2013-003, 2013 (unpublished). http://cds.cern.ch/record/1520896/files/CERN-ACC-NOTE-2013-003.pdf?subformat=pdfa.
[46] CERN, EDMS No. 875609, 2008 (unpublished).
[47] CERN, EDMS No 875606, 2008 (unpublished).
[48] L. Alberty, EDMS No. 1284428, 2013 (unpublished).
[49] B. P. Xiao, L. Alberty, S. Belomestnykh, *et al.*, in *Proceedings of the 4th International Particle Accelerator Conference, Shanghai, China, 2013,* p. 2417. http://accelconf.web.cern.ch/AccelConf/IPAC2013/papers/wepwo050.pdf.
[50] K. M. Wilson, E. F. Daly, J. Henry, *et al.*, in *Proceedings of the 2003 Particle Accelerator Conference, Portland, Oregon, 2003,* p. 2866. http://accelconf.web.cern.ch/AccelConf/p03/PAPERS/RPAB064.PDF.
[51] H. Padamsee, J. Knobloch, T. Hays, *RF Superconductivity for Accelerators*, John Wiley & Sons, (1998).

[52] K. Ko, A. Candel, L. Ge, *et al.*, in *Linear Accelerator Conference 2010, Tsukuba, Japan, 2010,* p. 1028. http://accelconf.web.cern.ch/accelconf/LINAC2010/papers/fr101.pdf.
[53] C. Marques, Master's thesis, Stony Brook University, (2014)
[54] G. Ciovati, Jefferson Lab Tech Note 03-003, 2003 http://tnweb.jlab.org/tn/2003/03-003.pdf.
[55] S. Belomestnykh, I. Ben-Zvi, C. Brutus, *et al.*, in *Proceedings of the 3rd International Particle Accelerator Conference, New Orleans, Louisiana, USA, 2012,* p. 2474. http://accelconf.web.cern.ch/accelconf/IPAC2012/papers/weppc109.pdf.
[56] Q. Wu, S. Belomestnykh, I. Ben-Zvi, in *Proceedings of 15th International Conference on RF Superconductivity, Chicago, Illinois, 2011,* p. 707. http://accelconf.web.cern.ch/AccelConf/SRF2011/papers/thpo007.pdf.